\documentclass[3p,review,pdftex,authoryear]{elsarticle}

\usepackage[T1]{fontenc}
\usepackage[utf8]{inputenc}
\usepackage{array}
\usepackage{amssymb, amsmath, amsthm}
\usepackage{graphicx}
\usepackage{lmodern,url}
\usepackage{makecell} % Sirve para hacer doble línea en una celda de tabla
\usepackage{cancel}
\usepackage{multirow}
\usepackage{microtype}
\usepackage{lineno}
\usepackage{xspace}
\usepackage{xcolor}
\usepackage{siunitx}
\usepackage{todonotes}
\usepackage{booktabs}
\usepackage[ruled,vlined]{algorithm2e}

\usepackage{mathdots}
\usepackage{subcaption}
\usepackage{dsfont}

\newcommand{\R}{\mathbb{R}}
\newcommand{\N}{\mathbb{N}}
\newcommand{\Prob}{\mathbb{P}}

\newcommand{\eg}{\emph{e.g.}\xspace}

\newcommand{\dis}{\displaystyle}

\newcommand{\tmin}{t_{\text{min}}}
\newcommand{\tmax}{t_{\text{max}}}


\definecolor{cfi}{rgb}{0,0.2,0.6}
\definecolor{beaublue}{rgb}{0.74, 0.83, 0.9}

\newcommand{\Ri}[1]{#1}
\begin{document}
%
%\linenumbers

\begin{frontmatter}
\title{Statistically-based methodology for revealing real contagion trends and correcting delay-induced errors in the assessment of COVID-19 pandemic}

\author[LRF,CEBIB]{Sebastián Contreras\fnref{now}\corref{cor}}
\ead{\Ri{sebastian.contreras@ds.mpg.de}}
\author[CEBIB,DIQBM]{Juan Pablo Biron-Lattes}
\author[CEBIB]{H. Andr\'es Villavicencio}
\author[CEBIB,CAL]{David Medina-Ortiz}
\author[CEBIB,DIQBM]{Nyna~Llanovarced-Kawles}
\author[CEBIB,DIQBM]{\'Alvaro Olivera-Nappa\corref{cor}}
\ead{aolivera@ing.uchile.cl}

\cortext[cor]{Corresponding Authors}
\fntext[now]{\Ri{Present address: Max Planck Institute for Dynamics and Self-Organization, Am Faßberg 17, 37077 Göttingen, Germany.}}
\address[LRF]{Laboratory for Rheology and Fluid Dynamics, Universidad de Chile, Beauchef 850, 8370448 Santiago, Chile.}
\address[CEBIB]{Centre for Biotechnology and Bioengineering, Universidad de Chile, Beauchef 851, 8370448 Santiago, Chile.}
\address[DIQBM]{Department of Chemical Engineering, Biotechnology, and Materials, Universidad de Chile, Beauchef 851, 8370448 Santiago, Chile.}
\address[CAL]{Division of Chemistry and Chemical Engineering, California Institute of Technology, Pasadena, CA 91125, USA}
\date{\today}
% Abstracts de ~100 palabras (no hay restricción). Paper sin estructura definida para Short communication ELSEVIER, with \leq than 6 Figures/Tables and \leq 1500 w.
\begin{abstract} 

COVID-19 pandemic has reshaped our world in a timescale much shorter than what we can understand. Particularities of SARS-CoV-2, such as its persistence in surfaces and the lack of a curative treatment or vaccine against COVID-19, have pushed authorities to apply restrictive policies to control its spreading. As data drove most of the decisions made in this global contingency, their quality is a critical variable for decision-making actors, and therefore should be carefully curated. In this work, we analyze the sources of error in typically reported epidemiological variables and usual tests used for diagnosis, and their impact on our understanding of COVID-19 spreading dynamics. We address the existence of different delays in the report of new cases, induced by the incubation time of the virus and testing-diagnosis time gaps, and other error sources related to the sensitivity/specificity of the tests used to diagnose COVID-19. Using a statistically-based algorithm, we perform a temporal reclassification of cases to avoid delay-induced errors, building up new epidemiologic curves centered in the day where the contagion effectively occurred. We also statistically enhance the robustness behind the discharge/recovery clinical criteria in the absence of a direct test, which is typically the case of non-first world countries, where the limited testing capabilities are fully dedicated to the evaluation of new cases. Finally, we applied our methodology to assess the evolution of the pandemic in Chile through the \Ri{Effective} Reproduction Number \Ri{$R_t$}, identifying different moments in which data was misleading governmental actions. In doing so, we aim to raise public awareness of the need for proper data reporting and processing protocols for epidemiological modelling and predictions. 

\end{abstract}

\end{frontmatter}

\section{Introduction}

Since the outbreak of novel SARS-CoV-2 in late 2019, the spread of COVID-19 has changed nearly every aspect of our daily life, challenging modern society to find a way to function under conditions never seen before. Governmental plans on public health have played a crucial role in its control in the absence of an effective treatment to cure COVID-19 or a vaccine to prevent it. Typically reported variables in the COVID-19 pandemic, available in public repositories \citep[as][among others]{Worldometers, dong2020interactive}, are the \Ri{incidence of new cases $\Delta T$, total cases $T$,} discharged/recovered cases $R$, and deaths $D$. These variables serve as input for the development and evaluation of governmental plans and to fit the vast variety of SIR-like mathematical models recently proposed \citep[see, \eg,][and references therein for a brief review of them]{contreras2020multigroup,yang2020modified,pang2020public,aguilar2020investigating}. Among the several factors conditioning the quality/reliability of the variables mentioned above are those related to the sensitivity and specificity of diagnostic tests, delays between sampling and diagnosing, and the delay between contagion, development of symptoms and testing. The latter varies from country to country, depending heavily on the local government's testing strategy and resources. Even though SIR-like models have been reported to be not-suitable to predict turning points or any other quantitative insight on this pandemic \citep{castro2020predictability}, to the best knowledge of the authors, delays have not been pointed out as the cause, as raw data has always been used to fit them.

Different parameters can be used to evaluate the evolution of the SARS-CoV-2 outbreak. Among them we may find the documentation rate \citep{wilder2020role}, secondary infection rate, serological response to infection \citep{world2020protocol}, number of vacancies at ICU \citep{liew2020preparing}, and the Basic Reproduction Number $R_0$ \citep{allieta2020covid, gevertz2020novel}, which is one of the most widely used.  This parameter ($R_0$) represents the number of persons a single infected individual might infect before either recovering or dying\Ri{, when the population is entirely susceptible} \citep{perasso2018introduction}. Traditional forms to estimate the $R_0$ are rather complex, heavily depending on the fitting of SIR models to local data \citep{heesterbeek2002brief,delamater2019complexity,PMID:32125128,MA2020129}.  In a previous work \citep{Contreras2020R0}, we proposed a methodology to obtain real-time estimations of \Ri{the Effective Reproduction Number $R_t$} directly from raw data, which was satisfactorily applied to evaluate the panorama of the COVID-19 spread in different countries and to forecast its evolution \citep{Medinaortiz2020Forecast}. Nevertheless, its heavy dependence on reported data required the study of common error sources affecting this parameter, and the development of methodologies to control, correct, and quantify their impact \citep{Contreras2020R0}.

In this work, we analyze the sources of error in the typically reported epidemiological variables and their impact on our understanding of COVID-19 spreading dynamics. We address the existence of different delays in the report of new cases, induced by the incubation time of the virus and testing-diagnosis time gaps, and provide a straightforward methodology to avoid the propagation of delay-induced errors to model-derived parameters. Using our statistically-based algorithm, we perform a temporal reclassification of individuals to the day where they were -statistically- most likely to have acquired the virus, building a new smooth curve with corrected variables. We postulate this new temporally modified curve as the proper one for fitting purposes in SIR-like models. We present an analogous methodology to estimate the number of discharged/recovered individuals, based on the reported evolution of the viral infection, the performance of the different tests for its diagnosis, and the case fatality, which can be easily adapted for a particular country. We used our methodology to assess the evolution of the pandemic in Chile, identifying different moments in which the use of raw data was misleading governmental actions. 

\section{On the performance of tests for diagnosing COVID-19}\label{STests}

Different methods for diagnosing COVID-19 have been developed and reported in the literature, with real-time RT-PCR being the standard applied globally \citep{lippi2020potential}. Nevertheless, techniques such as the IgG and IgM rapid tests, the Chest Computed Tomography (Chest CT), and CRISPR-Cas systems are also being used. In this section we provide a brief analysis of them, highlighting the different characteristics of both the techniques and their basic approach to detect viral infection.

\subsection{Real-time RT-PCR}

 Real-time reverse transcription polymerase chain reaction (RT-PCR) is a mechanism for amplification and detection of RNA in real time \citep[see][for an exhaustive description of the technique]{gibson1996novel}. Initially, RNA obtained from samples is retrotranscribed to DNA using a reverse transcriptase enzyme. By applying temperature cycles, the conditions are created for new copies of the DNA to be synthesized from the initial one. The lower the initial DNA concentration, the lower the probability of a synthesis reaction in a given cycle.
It is assumed that the minimum time from contagion until testing positive in the RT-PCR test is \SI{2.3}{days} (\SI{95}{\%} CI, \num{0.8}–\SI{3.0}{days}) before the onset of symptoms \citep{he2020temporal}, which typically appear 5.2 days (\SI{95}{\%} CI, \num{4.1}–\SI{7.0}{days}) after contagion \citep{lauer2020incubation}. \citet{he2020temporal,ganyani2020estimating} showed that \SI{44}{\%} to \SI{62}{\%} of the total contagions occur in the pre-symptomatic period. It has been inferred that the viral load reaches a peak value before 0.7 days from the onset of symptoms (\SI{95}{\%} CI, \num{0.2}–\SI{2.0}{days}), when it starts falling monotonically together with the infectivity rate \citep{he2020temporal}.  Finally, the virus has been detected for a median of 20 days after the onset of symptoms \citep{zhou2020clinical}, but infectivity may decrease significantly eight days after symptom appearance. 

A high false-negative rate \citep{li2020stability,xiao2020false} and a sensitivity of 71 to~\SI{83}{\%} \citep{fang2020sensitivity,long2020diagnosis} have been reported for the real-time RT-PCR technique, and several vulnerabilities of it have been identified and quantified \citep{lippi2020potential}. Considering sampling, handling, testing, and reporting, the total time necessary to get RT-PCR results for an individual may range between 2 to 3 days \citep{nguyen20202019}. However, the time it takes to perform the RT-PCR experiment takes about 2 to 3 hours \citep{li2020development}.

\subsection{IgG and IgM rapid tests}

Part of the immune system response to the SARS-CoV-2 infection is the production of specific antibodies against it, including IgG and IgM \citep{li2020molecular}. Serological tests detect the presence of those antibodies and, unlike the other detection methods, take only \SI{15}{minutes} to produce results \citep{li2020development}. These tests have a sensitivity of \SI{88.6}{\%} and a specificity of \SI{90.6}{\%} \citep{li2020development}. This technology was developed for the SARS-CoV epidemic, which was caused by a virus belonging to the same family of coronaviruses as SARS-CoV-2, providing satisfactory results after 2--3 days from the onset of symptoms (for IgG), and after eight days (for IgM) \citep{woo2004longitudinal}.

\subsection{Chest Computed Tomography}

The principle behind the Chest Computed Tomography (Chest CT) is the analysis of cross-sectional lung images to identify viral pneumonia features, like ground-glass opacity, consolidation, reticulation/thickened interlobular septa or nodules \citep{ai2020correlation}. Chest CT has shown a sensitivity between \SI{97}{\%} and~\SI{98}{\%} \citep{long2020diagnosis,fang2020sensitivity}. However, due to the similarities between CT images accounting for COVID-19 and CT images for other viral types of pneumonia, false-positives are likely to occur. Compared to RT-PCR, Chest CT tends to be more reliable, practical, and quick to diagnose COVID-19 \citep{ai2020correlation}. Nevertheless, requiring the presence of the potentially infected patient in a health center lacks the flexibility that rapid tests provide, and can backfire on movement restriction measures. COVID-19 pneumonia manifests with abnormalities on computed tomography images of the chest, even in asymptomatic patients \citep{shi2020radiological}.

\subsection{CRISPR-Cas systems}

In CRISPR-Cas systems, a guide RNA (gRNA) is designed to recognize a specific RNA sequence, like any particular gene or partial ARN sequence of SARS-CoV-2 coronavirus. Endonuclease enzymes of the Cas family and the specific gRNA will search for the sequence match. This match will deliver a signal that confirms the presence of SARS-CoV-2 RNA in the sample \citep{lemieux2020covid}. DETECTR and SHERLOCK are two examples of CRISPR-Cas technologies for the detection of SARS-CoV-2, being able to obtain results in less than 1 hour at a significantly lower cost compared to the RT-PCR technique. DETECTR showed a 95\% positive predictive agreement and 100\% negative predictive agreement \citep{broughton2020crispr}, while SHERLOCK has not been validated using real patient samples and is not suitable for clinical use at this time \citep{jin2020virology}.

\section{Methodology}

Our work aims to expose and quantify, both theoretically and in a case study, the impact of different sources of error in commonly reported data of the COVID-19 spread, such as newly reported cases $\Delta T$, total cases $T$, infected $I$, and recovered $R$ fractions of the population. 

First, we define random variables associated with the delay in both sampling and diagnosing new cases, and by modeling their probability distribution functions, we derive a method to re-classify the newly reported cases accordingly. As reclassification occurs backwardly, for re-evaluating the current scenario through our methodology, we cast predictions on the reported new cases using an ARIMA (Autoregressive Integrated Moving Average) model. Having the corrected variables, we evaluate differences on \Ri{the effective reproduction number $R_t$}, following the methodology presented by \citet{Contreras2020R0}\Ri{. Assuming the outbreak is well represented with a SIR model in a given timeframe $i$, $R_t$ can be estimated as $\frac{\beta_i}{\gamma_i}$, which can be directly estimated from the discrete version of the SIR differential equations. Decoupling the ``recovered'' fraction in the SIR model as clinically recovered individuals $R$ and deaths $D$, $R_t$ would be given by equation ~\ref{eqR0}:}
\begin{equation}\label{eqR0}
\Ri{R_t} = \frac{\Delta I}{\Delta R+ \Delta D}+1. 
\end{equation}
\Ri{Given the existence of a latent and incubation period for COVID-19, which will be carefully described in the next section, SIR models might not capture all the features of the outbreak and SEIR models would be more suitable. In such models, only infected individuals can propagate the infection and, equation-wise, estimations of $R_t$ remain more or less the same \citep{chowell2007comparative}. As data of exposed individuals is not widely available or entirely reliable, its contribution was not included in the estimations of $R_t$ to keep its straightforward use.}

Autoregression models for the forecast of $\Delta T$ were implemented using the \texttt{statsmodels} Python library \citep{seabold2010statsmodels}. All other calculations and visualizations were made using \texttt{DMAKit-lib} Python library \citep{dmakit} and MATLAB R2018a.

\section{Results}

\subsection{Temporal misclassification of new cases}

There exists an incubation period $t_c$ for the development of SARS-CoV-2-related symptoms, which average has been reported to range between 5.\Ri{1} days \citep{li2020early, lauer2020qifang,he2020temporal} and 6.4 days \citep{backer2020incubation}. \Ri{\citet{he2020temporal} also reported the existence of a latency period, which is the time required before an infected individual could spread the infection, which was extended until 2.3 days (95\% CI, \SI{0.8}{}–\SI{3.0}{days}) before the onset of symptoms, which occurred $t_c$ days after contagion.}

\begin{figure}[ht!]
    \centering
    \includegraphics[scale=.15]{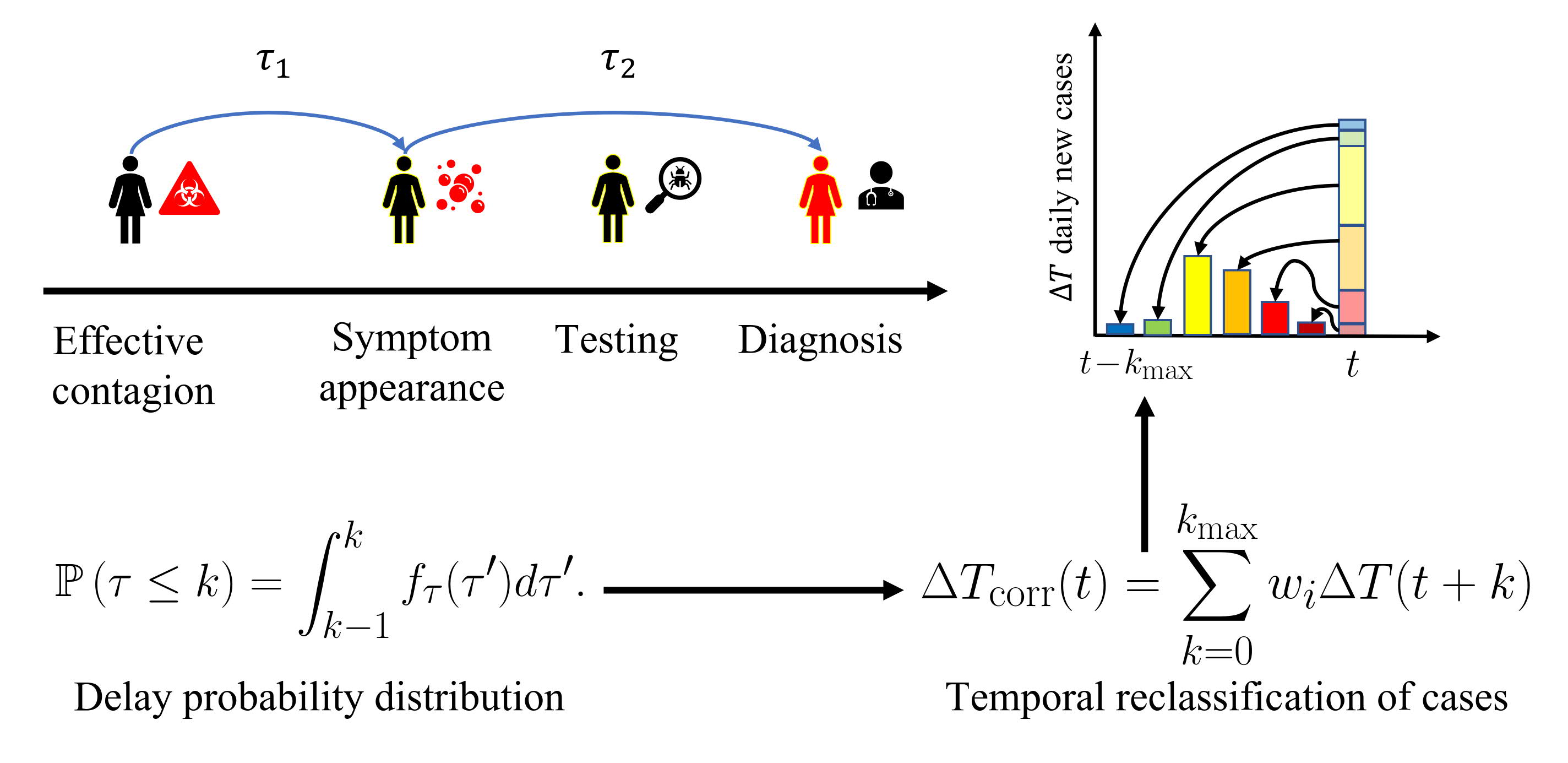}
    \caption{Schematic representation of the temporal reclassification methodology proposed herein.}
    \label{GraphAbstract}
\end{figure}

\Ri{The} incubation time $t_c$ is especially relevant in the case of a symptoms-based testing strategy or when the spread has reached the non-traceability stage. \Ri{To model it, we use} $\tau_1$ with \Ri{log-normal} distribution\Ri{, as reported in \citet{lauer2020incubation} and suggested in \citet{nishiura2007early}:}
\Ri{\begin{equation}\label{tau1}
    \tau_1 \sim \text{log}\,\mathcal{N}\left(\mu,\sigma^2\right).
\end{equation}}
\Ri{According to the values reported in \citet{lauer2020incubation}, $\mu=1.63=\ln(t_c)$ and $\sigma=0.41$ retrieve the expected values of 5.1 for the median, and 11.5 for the 98'th percentile. Even} though the required time for performing the test is short \citep{li2020development}, delays between testing and diagnosis have been reported  \citep{nguyen20202019}. We will sum up secondary delays, such as the symptom-to-testing and testing-to-diagnosis time gaps, into a random variable $\tau_2$, which, for the sake of simplicity, will be assumed to follow a uniform distribution between $\tmin$ and $\tmax$:
\begin{equation}\label{tau2}
\tau_2 \sim \mathcal{U}\left([\tmin,\tmax]\right).
\end{equation}
Further knowledge on the nature of factors affecting $\tau_2$ could lead to a different distribution. We may postulate a reclassification for obtaining the real number of new contagions occurred in a day $t$ as the sum of contributions of cases reported with a delay of $k$ days:
\begin{equation}\label{formulE}
    \Delta T_{\text{corr}} (t) = \sum_{k\in\N\cup\{0\}} w_i\Delta T(t+k),\qquad  k = \tau_1+\tau_2,
\end{equation}
where $w_i$ represents the fraction of patients that were notified at time $t' = t+k $ but had acquired the virus at $t$. Note that the different delays are referred to the random variable $Z=  \tau_1+\tau_2$. The probability distribution function for $Z$ is obtained by the convolution method, assuming $\tau_1$ and $\tau_2$ are independent and combining equations~\ref{tau1} and~\ref{tau2}:
\Ri{\begin{equation}\label{distZ}
    f_{Z}(z) = \left\{ \begin{array}{ll} 0, & \text{if }  z<\tmin \\
               \dis\frac{1}{2\left(\tmax-\tmin\right)}\left[1+\text{erf}\left(\frac{\ln\left(\dis\frac{z-\tmin}{t_c}\right)}{\sigma\sqrt{2}}\right)\right], & \text{if }  \tmin\leq z\leq\tmax \\ & \\
               \dis\frac{1}{2\left(\tmax-\tmin\right)}\left[\text{erf}\left(\frac{\ln\left(\dis\frac{z-\tmin}{t_c}\right)}{\sigma\sqrt{2}}\right)-\text{erf}\left(\frac{\ln\left(\dis\frac{z-\tmax}{t_c}\right)}{\sigma\sqrt{2}}\right)\right], & \text{if }  z> \tmax. \end{array} \right.
\end{equation}}
Assuming that data is reported on a daily basis, we can calculate the probability associated to having a delay of $k$ days:
\begin{equation}\label{masaZ}
    p_k  = \int_{k-1}^{k}f_{Z}(z)dz.
\end{equation}
For practical reasons, we can define a threshold $U<1$ for truncating the probability mass distribution (equation~\ref{masaZ}), which otherwise would assign a probability to every $k\in\N$. Let $n_1$ be the first positive integer for which equation~\ref{eqnUmbral} holds,
\begin{equation}\label{eqnUmbral}
    \Prob\left(Z\leq n_1\right)\geq U,
\end{equation}
we may rewrite equation~\ref{formulE} as:
\begin{equation}
    \Delta T_{\text{corr}} (t) = \frac{\dis\sum_{k=0}^{n_1} p_k\,\Delta T(t+k)}{\dis\sum_{k=0}^{n_1}p_k}.
\end{equation}
The magnitude of the total delay between infection and diagnosis can be estimated through the expected value of equation~\ref{distZ} (or equivalently, equation~\ref{masaZ}). A schematic representation of the proposed methodology is presented in Figure~\ref{GraphAbstract}. Assuming the lowest reported value for the average incubation time, $t_c = 5.2$, and a conservative timeframe for the delay between the appearance of symptoms, testing and diagnosing, (2--5 days), the expected delay $\mathbb{E}\left(Z\right)$ is about $\SI{7.5}{days}$.

\begin{algorithm}[!htp]
\SetAlgoLined
\KwResult{Corrected differential and cumulative total cases $T$.}
 $dT:$ Officially reported new cases per day. $dT\in\R^{n}$\;
 $\lambda, a,b:$  Parameters of the $\tau_1$ and $\tau_2$ distributions\;
 $U:$  Probability threshold for the maximum possible delay\;
Calculate $[Pk,F]$, respectively the probability and cumulative distribution functions for $Z=\tau_1+\tau_2$\;
 idx = $F$< $U$; logical indexes of the values of vector $F$ smaller than $U$\;
 $Pk$ = $Pk$(idx); $Pk\in\R^{m}$\;
 Define $K$, delay indexes of $Pk$ in idx. Renormalize $Pk$\;
 Forecast the next $m$ values for $dT$, using an autoregression (ARIMA) model\;
 Define an empty vector $dTcorr\in\R^{2m+n-1}$ to record the corrections\;
 \For{i = 1:\texttt{length}(dT) \do}{
$dTi = dT(i)$\;
$ind = m+i-K$\;
$dTcorr(ind) = dTcorr(ind) + dTi\cdot Pk$\;
 }
 Delete the last $m$ (incomplete) values from $dTcorr$; $dTcorr(end-m+1:end) = [\,]$\;
 Round $dTcorr$ values to the nearest integer; $dTcorr = round(dTcorr)$\;
 \caption{Statistically-based temporal reclassification algorithm for correcting delay-induced errors in the report of new COVID-19 infections}\label{CorrecciondT}
\end{algorithm}

\subsection{Case discharge/recovery criteria}

As discussed previously, errors in the amount of discharged/recovered patients $R$ are likely to be greater only when no quantitative criteria are applied. In such cases, some countries (like Chile) have adopted the following criteria (officially reported in the \cite{minsal} report), possibly based on the recommendations published by the \citet{WHO-quarantine}.

\begin{itemize}
	\item If there were no previously existing pathologies, a patient should be discharged 14 days after testing positive for COVID-19.
	\item If there were previously existing pathologies, the patient should be discharged 28 days after testing positive for COVID-19.
\end{itemize}
This criterion turns out to be quite simplistic, especially considering the existence of uncertainties regarding the diagnosis and contagion days. If we try to model the probability of recovering from COVID-19, some assumptions are necessary. Let $\tau_r$ be the random variable for the time of discharge/recovery such that:

\begin{itemize}
    \item $\mathbb{P}\left(\tau_r\leq 14\right)= 0$
    \item $\mathbb{P}\left(\tau_r\leq 28\right)\approx 1$
\end{itemize}

Further assumptions are necessary to estimate the probability distribution function $f_{\tau_r}$, as it depends on local diagnosis criteria, testing strategy, and the fraction of the population having preexisting pathologies. In particular, depending on the test applied for diagnosis and its sensitivity/specificity -- which were carefully described in Section~\ref{STests}-- the probability profile would change. The simplest form that can be assumed, and which, for clarity reasons, is adopted herein, is a triangular distribution:
\begin{equation}
    f_{\tau_r}(t) = \left\{ \begin{array}{ll} 0, & \text{if }  t<\tmin' \\ \dis\frac{2}{\tmax'-\tmin'}\left(1-\frac{t-\tmin'}{\tmax'-\tmin'}\right), & \text{if }  \tmin'\leq t \leq \tmax' \\ 0, & \text{if }  t>\tmax'. \end{array} \right.
\end{equation}
Further knowledge on the nature of factors affecting $\tau_r$ could lead to a different distribution. 
\begin{algorithm}[!htp]
\SetAlgoLined
\KwResult{Estimated discharged/recovered cases per day $R$.}
 $dTcorr:$ corrected new cases per day. $dTcorr\in\R^{n}$\;
 $dD:$ Officially reported daily deaths due to COVID-19. $dD\in\R^{n}$\;
 $\vec{\theta}:$  Parameters of the probability distribution function for the time of discharge/recovery $\tau_r$.\;
 $\tmin,\,\tmax:$ time frame for discharging an infected patient.\;
 $w = \tmax\!-\!\tmin\!+\!1$, width of the time-frame of discharge\;
 $K = \texttt{linspace}(\tmin,\tmax,w)$\;
Calculate the probability distribution function for $\tau_r$; $PkR = f(\vec{\theta},w)$\;
 Define an empty vector $dRcorr\in\R^{w+n}$ to record the corrections\;
 \For{i = 1:n-$\tmin$ \do}{
$dTi = dTcorr(i)$\;
$ind = n+i-K$\;
$dRcorr(ind) = dRcorr(ind) + dTi\cdot PkR$;
 }
 Delete the last $w$ (incomplete) values from $dRcorr$; $dRcorr(end-(w-1):end) = [\,]$\;
 Subtract the daily deaths due to COVID-19 to obtain the final estimator, round to the nearest integer. Confirm the result is positive or zero.\;
 $dRcorr = \max (\text{round}(dRcorr-dD),0)$ \;
 \caption{Statistical estimation of $dR$, daily patients that have been discharged/recovered from COVID-19}\label{EstimaciondR}
\end{algorithm}

\section{Case study: COVID-19 spreading dynamics in Chile}

\begin{figure}[ht!]
    \centering
    \includegraphics[scale=.45]{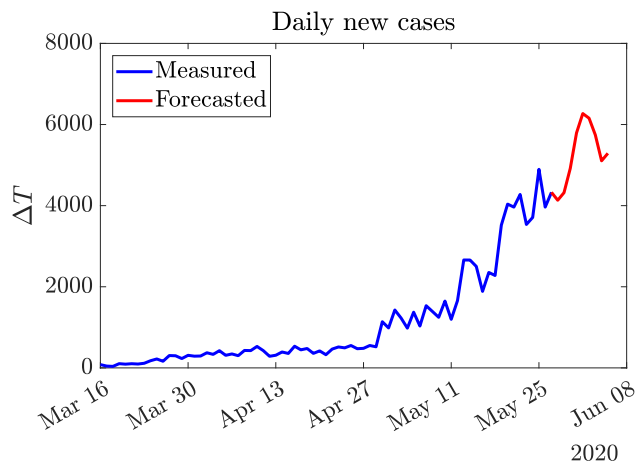}
    \caption{Current and forecasted evolution of $\Delta T$ in Chile. The official data (blue curve) was obtained from \cite{Worldometers}, while the red curve was generated using an autoregression ARIMA model. \Ri{As these reported data are likely to be affected by several exogenous factors \citep{ribeiro2020short}, the best performance metric for the generated forecast is trend consistency.}}
    \label{FigForecast}
\end{figure}

The spread of COVID-19 in Chile is far from being controlled, as shown by the exponential growth that new cases have had in recent weeks \citep{Worldometers}. In order to apply our methodology, we need to cast predictions on the trends of $\Delta T$. Figure~\ref{FigForecast} presents the current and forecast trends, using an autoregression ARIMA model. \Ri{These models were chosen over others because of the great success they have had on representing COVID-19 dynamics in other relevant studies \citep{chakraborty2020real,benvenuto2020application}.}

\begin{figure}[ht!]
    \centering
    \includegraphics[scale=.45]{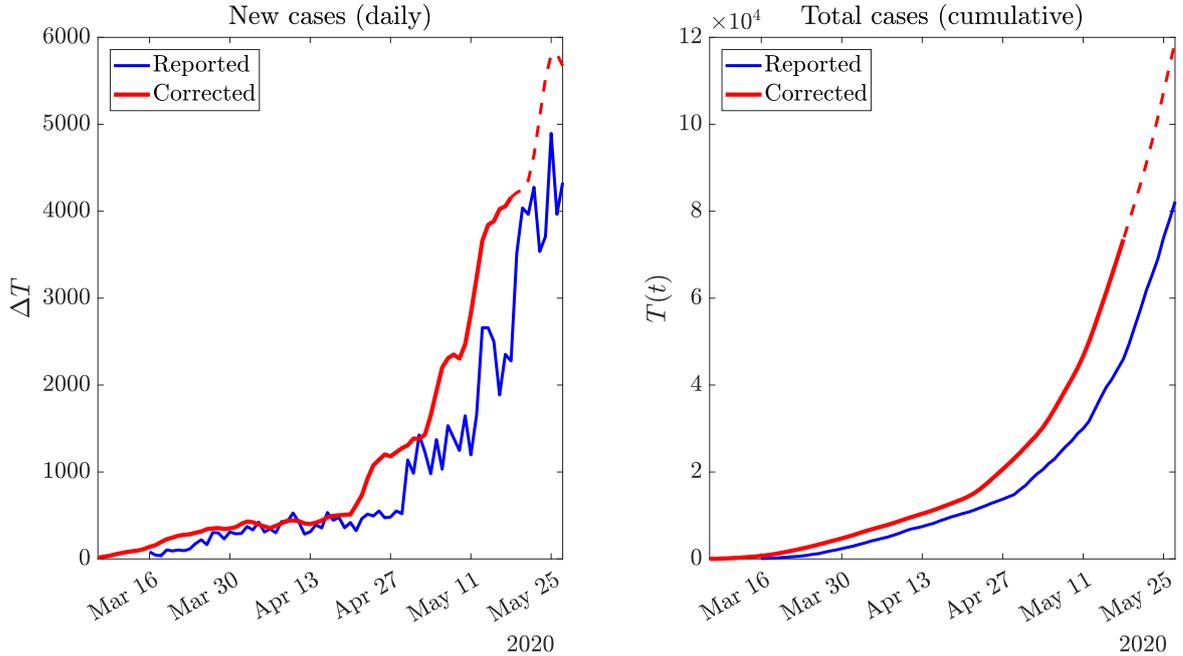}
    \caption{Statistically-driven reclassification of new cases (red curve) compared with the raw data (blue curve), both differential a) and cumulative b). The last dashed part of the red curve accounts for the values which depends in the forecast shown in Figure~\ref{FigForecast}, and therefore might change over time. Official data obtained from \cite{Worldometers}.}
    \label{FigCorr}
\end{figure}

First, we performed a temporal reclassification of new cases to obtain $\Delta T_{\text{corr}}$, presented in Figure~\ref{FigCorr}. It can be seen that our methodology, besides exhibiting an horizontal semi-displacement, generates a smooth curve, which is due to the inclusion of a probability function that spreads the influence of daily informed cases in previous days (where they were most likely to have occurred). The last part of the red curve is dashed because it partially contains contributions of the forecast of $\Delta T$, and therefore might change in the upcoming days, when the required data for completing the reclassification of cases would be available.

After obtaining statistically-based corrections of both $\Delta T$ and $\Delta R$ by following algorithms~\ref{CorrecciondT} and~\ref{EstimaciondR}, and knowing the daily deaths due to COVID-19 $\Delta D$, we proceed to calculate the variation on the active cases. \Ri{Based on the closure of the population-balance of the different considered classes, the cumulative new cases $T$ will be the sum of active $I$, recovered/discharged $R$, and dead $D$, therefore their differences would follow:}
\Ri{\begin{equation}\label{eqdi}
    \Delta T = \Delta I + \Delta R + \Delta D \Longleftrightarrow \Delta I = \Delta T-(\Delta R + \Delta D).
\end{equation}}
\Ri{In the context of SEIR models, it is not possible to explicitly include exposed individuals $E$, as they are not being recorded separately from COVID-19 positive new cases.}

\begin{figure}[ht!]
    \centering
    \includegraphics[scale=.4]{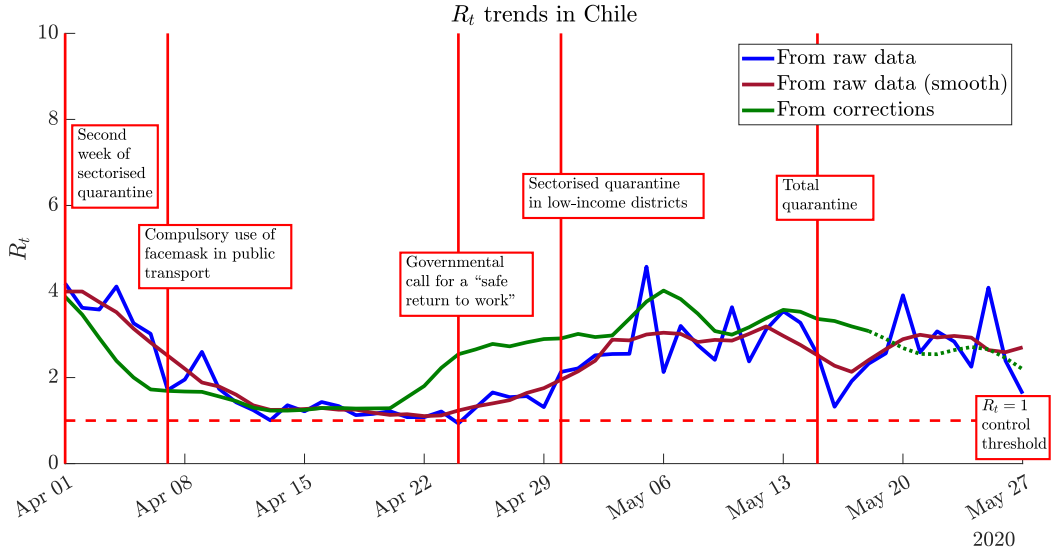}
    \caption{Effect of data processing on the evaluation of the spread of COVID-19 through \Ri{$R_t$}. The noisy raw data (blue curve) can be smoothed through mobile averages (dark red curve), but the trends are the same. A significantly different scenario is shown by the statistically-corrected \Ri{$R_t$} trend (green curve). Highlighted dates associated with iconic governmental actions in Chile: March 31st (second week of sectorised quarantine for the high-income districts of Santiago, capital of Chile), April 7th (compulsory use of facemask in public transport), April 24th (governmental call for a ``safe return to work''), April 30th (sectorised quarantine --low-income districts of Santiago--), May 15th (total quarantine in Santiago).}
    \label{R0Chilito}
\end{figure}

\Ri{After obtaining $\Delta I$ using equation~\ref{eqdi},} we proceed to calculate \Ri{$R_t$}, using equation~\ref{eqR0} with raw data, mobile averages of raw data, and the methodology proposed herein. As shown in Figure~\ref{R0Chilito}, an abrupt growth in \Ri{$R_t$} was evidenced around April 22nd, consistently with the relaxation of restrictive measures that were applied in Santiago, the capital and most populated city in the country, and the apogee of the governmental plan for a ``safe return to work''. Even though different trends seem to stabilise again in the second week of May, a lowering of the trend is not totally clear.

The different iconic dates highlighted in Figure~\ref{R0Chilito} were obtained from the chronology presented in \cite{Wikipedia} and references therein. From this plot, we can assess both successful and unfortunate effects that the different governmental actions have had on the spreading dynamics of COVID-19 in Chile, and specifically in Santiago, by analyzing corrected \Ri{$R_t$} trends and comparing them with daily reported cases. The various public-health actions were strongly based on locally-observed values from (raw) daily reported cases, yet appear to have been taken too late, or too early, according to the statistically-corrected trend. In particular, the apogee of the governmental plan for a safe return to work occurred in a temporal window where raw-data-driven \Ri{$R_t$} values were at a minimum, but the corrected contagion rates anticipated a steep growing trend for the following days.

\section{Conclusions}

We have presented an exhaustive assessment of error sources in reported data of the COVID-19 pandemic and provided a methodology to minimize --and correct-- their effect on both reported variables and model-derived parameters by applying a statistically-driven reclassification of newly reported cases, and corrections to discharge/recovery criteria. By using the corrected variables, SIR-like models could be fitted directly, as every value would represent the real contagion dynamics. We present the methodology as a general framework, aiming to provide a useful tool for researchers and decision-making actors looking to adapt it for their particular interests.

In a case study on the spreading dynamics of COVID-19 in Chile, we observed the effects that different iconic public health actions taken by the government had on SARS-CoV-2 spread rate \Ri{$R_t$}, readily calculated using our previously reported approach, and discussed the reasons behind these effects interpreted under the light of raw data and corrected daily contagion rates, calculated with the methodology presented in this manuscript. The delay-induced error in raw data slowed-down the reaction time, so the actions were taken too late for restrictive actions or too early for the comeback to normality. Our statistically-driven method corrected such reporting errors, exposing the real contagion dynamics at a given time. The proposed methodology also serves as a non-invasive smoothing process, as it only temporally re-sorts daily reported cases according to their most likely report delay from the real contagion day.

We expect our methodology to serve as a valuable input for researchers and public health practitioners trying to add statistical value to their calculations and predictions in order to vanquish the current SARS-CoV-2 pandemic. Finally, we also expect this work to contribute in raising public awareness on the need for a proper (and standardized) strategy for report and curation of data in the COVID-19 pandemic and other highly-contagious diseases.

%\appendices

\section*{Conflict of Interest Statement}

The authors declare that the research was conducted in the absence of any commercial or financial relationships that could be construed as a potential conflict of interest.

\section*{CRediT author statement}

\textbf{Sebastián Contreras:} Conceptualization, Methodology, Investigation, Writing- Original draft preparation, Visualization, Writing- review and editing, Supervision.  \textbf{Juan Pablo Biron-Lattes:} Validation, Investigation, Writing- Original draft preparation, Writing- review and editing.  \textbf{H. Andrés Villavicencio:} Conceptualization, Investigation, Writing- Original draft preparation, Supervision.  \textbf{David Medina-Ortiz:} Validation, Writing- review and editing.  \textbf{Nyna Llanovarced-Kawles:} Investigation, Writing- Original draft preparation.  \textbf{Álvaro Olivera-Nappa:} Validation, Writing- review and editing, Supervision, Project administration, Funding resources.

\section*{Acknowledgements}
The authors gratefully acknowledge support from the Chilean National Agency for Research and development through ANID PIA Grant AFB180004, and the Centre for Biotechnology and Bioengineering - CeBiB (PIA project FB0001, Conicyt, Chile). DM-O gratefully acknowledges Conicyt, Chile, for PhD fellowship 21181435. 

%\appendix
%\section{Experimental protocol}\label{apendiceexp}

\section*{References}
\bibliographystyle{elsarticle-harv}
%\bibliography{document}

% that's all folks
\end{document}